\documentclass[11pt]{article}
\usepackage{amssymb}
\usepackage{amsmath}
\usepackage{graphicx,color}
\usepackage{wrapfig,framed}

\newtheorem{theorem}{Theorem}
\newtheorem{guess}[theorem]{Conjecture}

\newtheorem{prop}[theorem]{Proposition}

\newtheorem{remark}[theorem]{Remark}

\newcommand{\eqa}{\begin{eqnarray}}
\newcommand{\eeqa}{\end{eqnarray}}
\newcommand{\beq}{\begin{equation}}
\newcommand{\eeq}{\end{equation}}
\newcommand{\nn}{\nonumber}

\newcommand{\pal}{\partial}

\newcommand{\F}{{\cal F}}

\newcommand{\bt}{{\bf t}}

\newcommand{\tr}{{\rm tr}}

\newcommand{\epf}{$\quad$\hfill
\raisebox{0.11truecm}{\fbox{}}\par\vskip0.4truecm}

\usepackage{array}
\newcolumntype{M}[1]{>{\centering\arraybackslash}m{#1}}
\newcolumntype{R}[1]{>{\raggedleft\arraybackslash}m{#1}}
\newcolumntype{N}{@{}m{0pt}@{}}
\allowdisplaybreaks[1]
\usepackage{hyperref}
\hypersetup{
    colorlinks,
    citecolor=black,
    filecolor=black,
    linkcolor=black,
    urlcolor=black}

\begin{document}

\title{On cubic Hodge integrals and random matrices}
\author
{
{Boris Dubrovin, Di Yang}\\
{\small SISSA, via Bonomea 265, Trieste 34136, Italy}\\
}
\date{}
\maketitle
\begin{abstract} A conjectural relationship between the GUE partition function with even couplings and certain special cubic Hodge integrals over the moduli spaces of stable algebraic curves is under consideration.
\end{abstract}

\tableofcontents

\section{Introduction}
\setcounter{equation}{0}
\setcounter{theorem}{0}
\subsection{Cubic Hodge partition function}
Let $\overline{\mathcal{M}}_{g,k}$ denote the Deligne--Mumford moduli space of stable curves of genus $g$ with 
$k$ distinct marked points. Denote by $\mathcal{L}_i$ the $i^{th}$ tautological line bundle over $\overline{\mathcal{M}}_{g,k}$, 
and $\mathbb{E}_{g,k}$ the rank $g$ 
Hodge bundle. Let $\psi_i:=c_1(\mathcal{L}_i),\, i=1,\dots,k,$  and let $\lambda_i:= c_i(\mathbb{E}_{g,k}),\, i=0,\dots,g.$ 
Recall that the Hodge integrals over $\overline{\mathcal{M}}_{g,k}$, \textit{aka} the intersection numbers of $\psi$- and $\lambda$-classes, are integrals of the form
$$
\int_{\overline{\mathcal{M}}_{g,k}} \psi_1^{i_1}\cdots \psi_k^{i_k} \cdot \lambda_1^{j_1} \cdots \lambda_g^{j_g}, \qquad i_1,\dots,i_k,\, j_1,\dots,j_g \geq 0. 
$$
Note that the dimension-degree matching implies that the above integrals vanish unless
$$
3g-3+k=(i_1+i_2+\dots+i_k)+  (j_1+2\,j_2+3\, j_3+\dots+g \, j_g).
$$

The particular case of \emph{cubic Hodge integrals} of the form
\beq\label{pqr}
\int_{\overline{\mathcal{M}}_{g,k}}  \Lambda_g(p) \, \Lambda_g(q) \, \Lambda_g(r) \, \psi_1^{i_1}\cdots \psi_k^{i_k}, \qquad \quad \frac1{p}+\frac1{q}+\frac1{r}=0
\eeq
was intensively studied after the formulation of the celebrated R.\,Gopakumar--M.\,Mari\~no--C.\,Vafa conjecture \cite{GV, MV} regarding the Chern--Simons/string duality. Here we denote
$$
\Lambda_g(z)= \sum_{i=0}^g \lambda_i \, z^i
$$
the Chern polynomial of $\mathbb{E}_{g,k}$. A remarkable expression for the cubic Hodge integrals of the form
$$
\int_{\overline{\mathcal{M}}_{g,k}} \frac{\Lambda_g(p)\Lambda_g(q)\Lambda_g(r)}{(1-x_1 \, \psi_1)\dots (1-x_k \, \psi_k)}, \quad k\geq 0
$$
conjectured in \cite{MV}
was proven in \cite{LLZ1, OP}; for more about cubic Hodge integrals see in the subsequent papers \cite{LLZ2, LLZ3, Zhou, DLYZ}.

In the present paper we will deal with the specific case of Hodge integrals \eqref{pqr} with a pair of equal parameters among $p$, $q$, $r$; without loss of generality one can assume that $p=q=-1$, $r=1/2$. So, the {\it special
cubic Hodge integrals} of the form
\beq\label{chbcy}
\int_{\overline{\mathcal{M}}_{g,k}} \Lambda_g(-1)\,\Lambda_g(-1)\,\Lambda_g\left(\frac12\right)\,\psi_1^{i_1}\cdots \psi_k^{i_k}
\eeq
will be considered.
Denote
\begin{equation} \label{cubic-hodge}
\mathcal H(\bt;\epsilon)=\sum_{g\geq 0} \epsilon^{2g-2} \sum_{k\geq 0}  \frac 1{k!} 
\sum_{i_1, \dots, i_k\geq 0} t_{i_1}\cdots t_{i_k}\int_{\overline{\mathcal M}_{g,k}} 
\Lambda_g(-1)\,\Lambda_g(-1)\,\Lambda_g\left(\frac12\right)\, \psi_1^{i_1}\cdots \psi_k^{i_k}
\end{equation}
the generating function of these integrals.
Here and below $\bt=(t_0,t_1,\dots)$ are independent variables, $\epsilon$ is a parameter. The exponential $e^{\mathcal H}=:Z_{\mathbb{E}}$ is called the cubic 
Hodge partition function while $\mathcal H(\bt;\epsilon)$ is the cubic Hodge free energy. It can be written in the form of genus expansion
\beq\label{genusexpand}
\mathcal{H}(\bt;\epsilon)=\sum_{g\geq0} \epsilon^{2g-2} \mathcal{H}_g(\bt)
\eeq
where $\mathcal{H}_g(\bt)$ is called the genus $g$ part of the cubic Hodge free energy, $g\geq 0$.
Clearly ${\mathcal H}_0(\bt)$ coincides with the Witten--Kontsevich generating function of genus zero intersection numbers of $\psi$-classes
\beq\label{H0}
\mathcal{H}_0(\bt)= \sum_{k\geq 0}  \frac 1{k!} 
\sum_{i_1, \dots, i_k\geq 0} t_{i_1}\cdots t_{i_k}\int_{\overline{\mathcal M}_{0,k}}  \psi_1^{i_1}\cdots \psi_k^{i_k}=\sum_{k\geq 3} \frac1{k(k-1)(k-2)}\sum_{i_1+\cdots+ i_k=k-3} \frac{t_{i_1}}{i_1!}\cdots \frac{t_{i_k}}{i_k!}.
\eeq
We note that an efficient algorithm for computing $\mathcal{H}_g(\bt)$, $g\geq 1$ was recently proposed in \cite{DLYZ}.



\setcounter{equation}{0}
\setcounter{theorem}{0}
\subsection{GUE partition function with even couplings}
Let ${\mathcal H}(N)$ denote the space of $N\times N$ Hermitean matrices. Denote
$$
dM=\prod_{i=1}^N dM_{ii} \prod_{i<j} d{\rm Re} M_{ij}\,d{\rm Im}M_{ij}
$$
the standard unitary invariant volume element on  ${\mathcal H}(N)$. The most studied Hermitean random matrix model is governed by the following GUE partition function with even couplings
\beq\label{part}
Z_N({\bf s})=\frac{(2\pi)^{-{N}} }{Vol(N)} \int_{{\mathcal H}(N)} e^{- N\, \tr \,V(M; \, {\bf s})} dM.
\eeq
Here, $V(M; \, {\bf s})$ is an \textbf{even} polynomial of $M$
\beq\label{pot}
V(M;{\bf s})=\frac12 M^2 -\sum_{j\geq 1} s_j \,  M^{2j},
\eeq
or, more generally, a power series,  by ${\bf s}=(s_1,s_2,s_3,\dots)$ 
we denote the collection of 
coefficients\footnote{The notation here is slightly different from that of \cite{icmp,DY1} where 
the coefficient of $M^{2j}$ was denoted by $s_{2j}$.} 
of $V(M)$, 
and by $Vol(N)$ the volume of the quotient of the unitary group over the maximal torus $\left[ U(1)\right]^N$
\beq\label{vol}
Vol(N)=Vol\left( U(N)/\left[ U(1)\right]^N\right)=\frac{\pi^{\frac{N(N-1)}2} }{G(N+1)}, \qquad G(N+1)=\prod_{n=1}^{N-1} n!.
\eeq
The integral will be considered as a formal saddle point expansion with respect to the small parameter $\epsilon$. Introduce the \emph{'t Hooft coupling} parameter $x$ by
$$
x:=N\, \epsilon.
$$
Reexpanding the free energy $\mathcal{F}_N({\bf s}):=\log Z_N({\bf s})$ in powers of $\epsilon$ and replacing the Barnes $G$-function by its asymptotic expansion yields\footnote{It is often called $1/N$-expansion as $\epsilon\sim 1/N$.}
\beq\label{genusF}
\mathcal{F}(x, {\bf s}; \epsilon):=\mathcal{F}_N({\bf s}) |_{N=\frac{x}{\epsilon}} -\frac1{12}\log \epsilon=\sum_{g\geq 0} \epsilon^{2g-2} \mathcal{F}_g (x, {\bf s}).
\eeq
The GUE free energy $\mathcal{F}(x, {\bf s}; \epsilon)$ can be represented \cite{thooft1, thooft2, BIZ, mehta} in the form
\eqa
&&  \hspace{-6mm}
\mathcal{F}(x, {\bf s};\epsilon ) = \frac{x^2}{2\epsilon^2} \left( \log x -\frac32\right) -\frac1{12} \log x +\zeta'(-1) +\sum_{g\geq 2} \epsilon^{2g-2} \frac{B_{2g}}{4g(g-1)x^{2g-2}} \nn\\
&& \hspace{-6mm}
\qquad\qquad\quad\, + \sum_{g\geq 0} \epsilon^{2g-2} \sum_{k\geq 0} \sum_{i_1, \dots, i_k\geq 1} a_g(i_1, \dots, i_k) \, s_{i_1} \dots s_{i_k} \, x^{2-2g -\left(k-{|i|}\right)}, \label{expan} \\
&& \hspace{-6mm} a_g(i_1, \dots, i_k) =
 \sum_{\Gamma} \frac{1}{\#\, {\rm Sym} \, \Gamma}
\eeqa
where the last summation is taken over all connected oriented ribbon graphs $\Gamma$ of genus $g$ with $k$ vertices of valencies $2i_1$, \dots, $2i_k$, 
$\#\,{\rm Sym}\, \Gamma$ is the order of the symmetry group of $\Gamma$, and $|i|:=i_1+\dots + i_k.$

Our goal is to compare the expansions \eqref{cubic-hodge} and \eqref{expan}. 

\setcounter{equation}{0}
\setcounter{theorem}{0}
\subsection{From cubic Hodge integrals to random matrices. Main Conjecture.} 

It was already observed by E.\,Witten \cite{Witten} that the GUE partition function with an even polynomial $V(M)$ is tau-function of a particular solution to the Volterra (also called the discrete KdV) hierarchy. Recall that the first equation of the hierarchy (the Volterra lattice equation) reads
$$
\dot w_n=w_n\left(w_{n+1}-w_{n-1}\right)
$$
where
$$
w_n=\frac{Z_{n+1} Z_{n-1}}{Z_n^2},
$$
the time derivative is with respect to the variable $t=N\, s_1$.
Other couplings $s_k$ are identified with the time variables of higher flows of the hierarchy. On another side, the study \cite{DLYZ} of integrable systems associated with the Hodge integrals\footnote{The first example of an integrable system associated with \emph{linear} Hodge integrals was investigated by A.\,Buryak. In this case the integrable system was proved to be Miura equivalent to the Intermediate Long Wave equation \cite{Bur}.} suggested the following conjectural statement: the Hodge partition function $Z_{\mathbb E}=e^{\mathcal{H}}$ of the form \eqref{cubic-hodge} as function of independent parameters $t_i$ is also a tau-function of the Volterra hierarchy. This observation provides a motivation for the main conjecture of the present paper.

It will be convenient to change normalisation of the GUE couplings. Put
$$
\bar s_k:=\left(\begin{array}{c} 2k\\ k\\ \end{array}\right) s_{k}.
$$

\begin{guess} [Main Conjecture] \label{c1} 
The following formula holds true
\eqa
&& \hspace{-6mm} \sum_{g=0}^\infty \epsilon^{2g-2} \mathcal{F}_g(x, {\bf s}) + \epsilon^{-2} \Big(-
\frac12  \sum_{k_1,k_2\geq 1} \frac{k_1\,k_2}{k_1+k_2}  \,  \bar{s}_{k_1}  \, \bar{s}_{k_2} + \sum_{k\geq 1} \frac{k}{1+k}\, \bar{s}_{k} -
x \,  \sum_{k\geq 1} \,  \bar{s}_{k}   - \frac14  + x \Big) \nn\\
&& \hspace{-6mm}  \qquad\qquad\qquad= \cosh\left(\frac{\epsilon \, \partial_x}2\right) \, \left[\sum_{g=0}^\infty \epsilon^{2g-2} \, 2^g \, \mathcal{H}_g\left(\bt(x, {\bf s})\right)\right].
\eeqa
where 
\beq\label{time-sub}
t_i(x, {\bf s}) := \sum_{k\geq 1} k^{i+1} \bar{s}_k  - 1 + \delta_{i,1} + x \cdot \delta_{i,0},\qquad i\geq 0.
\eeq
\end{guess}

\setcounter{equation}{0}
\setcounter{theorem}{0}
\subsection{Computational aspects of the Main Conjecture: how do we verify it?}

We will check validity of the Main Conjecture for small genera. Begin with $g=0$. Let us start with $\mathcal{H}_0(\bt)$. Instead of the explicit expansion \eqref{H0} we use the following well known representation
\beq\label{h0}
\mathcal{H}_0=\frac{v^3}6-\sum_{i\geq 0}t_i \frac{v^{i+2}}{i! (i+2)}+\frac12\sum_{i, \, j\geq 0}  t_i  t_j \frac{v^{i+j+1}}{(i+j+1) \, i! \, j!}
\eeq
where $v=v({\bt})=t_0+\dots$ is the unique series solution to the equation
\beq\label{eqv}
v=\sum_{i\geq 0} t_i \frac{v^i}{i!}.
\eeq
Here we recall that
\beq\label{defv}
v=\frac{\pal^2 \mathcal{H}_0(\bt)}{\pal t_0^2}=\sum_{k=1}^\infty\frac1{k}\sum_{i_1+\dots+i_k=k-1}\frac{t_{i_1}}{i_1!}\dots \frac{t_{i_k}}{i_k!}
\eeq
is a particular solution to the Riemann--Hopf hierarchy
$$
\frac{\pal v}{\pal t_k}=\frac{v^k}{k!} \frac{\pal v}{\pal t_0}, \qquad k=0, \, 1, \, 2, \dots .
$$

For the genus zero GUE free energy $\mathcal{F}_0=\mathcal{F}_0(x,{\bf s})$ one has a similar representation. Like above, introduce
\beq\label{defu}
u(x,{\bf s})=\frac{\pal^2 \F_0(x,{\bf s})}{\pal x^2}
\eeq
and put
\beq\label{defw}
w(x,{\bf s})=e^{u(x,{\bf s})}.
\eeq

\begin{prop} \label{F0thm} 
The function $w=w(x,{\bf s})$ is the unique series solution to the equation
\beq\label{weven}
w= x+ \sum_{k\geq 1} \, k\, \bar s_k  \, w^k,\qquad \bar s_k:=\left(\begin{array}{c} 2k\\ k\\ \end{array}\right) s_{k}, \quad w(x,{\bf s})=x+\dots.
\eeq
The genus zero GUE free energy $\F_0$ with even couplings has the following expression
\beq \label{F0even}
\mathcal{F}_0 = \frac{w^2}4 - x \, w + \sum_{k\geq 1} \bar{s}_k \left(x \, w^k-\frac{k}{k+1} w^{k+1}\right) +\frac12 \sum_{k_1,k_2\geq 1} \frac{k_1k_2}{k_1+k_2} 
\bar{s}_{k_1} \bar{s}_{k_2} w^{k_1+k_2} + \frac{x^2} 2\, \log w.
\eeq
\end{prop}

Clearly $w$  also satisfies the Riemann--Hopf hierarchy in a different normalization
$$
\frac{\pal w}{\pal \bar s_k} =k\, w^k \frac{\pal w}{\pal x}, \qquad k\geq 1.
$$
The solution can be written explicitly in the form essentially equivalent to \eqref{defv}
$$
w=\sum_{n=1}^\infty \frac1{n} \sum_{i_1+\dots + i_{n}=n-1} {\rm wt}(i_1)\dots {\rm wt}(i_n) \, \bar s_{i_1}\dots \bar s_{i_n}
$$
where we put $\bar s_0=x$ and denote
$$
{\rm wt}(i)=\left\{ \begin{array}{cl} 1, & i=0\\
\\
i, & {\rm otherwise.}\end{array}\right. .
$$

It is now straightforward to verify that the substitution \eqref{time-sub} yields
\beq\label{vw}
e^{v\left( \bt(x,{\bf s})\right)}=w(x, {\bf s}), \qquad \mbox{i.e. } v\left( \bt(x,{\bf s})\right)=u(x,{\bf s})
\eeq
and
\beq\label{main0}
\mathcal{H}_0\left(\bt(x,{\bf s})\right)=\F_0\left(x,{\bf s}\right)-
\frac12  \sum_{k_1,k_2\geq 1} \frac{k_1\,k_2}{k_1+k_2}  \,  \bar{s}_{k_1}  \, \bar{s}_{k_2} + \sum_{k\geq 1} \frac{k}{1+k}\, \bar{s}_{k} -
x \,  \sum_{k\geq 1} \,  \bar{s}_{k}   - \frac14  + x.
\eeq
See in Sect.\,\ref{DZapproach} for the details of this computation.

\medskip

In order to proceed to higher genera we will use the method that goes back to the paper \cite{DW} by R.\,Dijkgraaf and E.\,Witten. The idea of this method is to express the positive genus free energy terms via the genus zero. Let us first explain this method for the Hodge free energy.

\begin{theorem}[\cite{DLYZ}] \label{polyHodge} There exist functions $H_g(v,v_1, v_2, \dots, v_{3g-2})$, $g\geq 1$ of independent variables $v$, $v_1$, $v_2$, $\dots$ such that
\beq\label{quasiH}
\mathcal{H}_g(\bt)=H_g\left( v(\bt ), \frac{\pal v(\bt )}{\pal t_0}, \dots, \frac{\pal^{3g-2} v(\bt)}{\pal t_0^{3g-2}}\right), \quad g\geq 1.
\eeq
Here $v({\bf t})$ is given by eq. \eqref{defv}.
Moreover, for any $g\geq 2$  the function $H_g$ is a polynomial in the variables $v_2$, \dots, $v_{3g-2}$ with coefficients in $\mathbb Q\left[ v_1, v_1^{-1}\right]$ (independent of $v$).
\end{theorem}

Explicitly,
\beq\label{H1}
H_1(v, v_1)=-\frac1{16} v+\frac1{24} \log v_1
\eeq
\beq\label{H2}
H_2(v_1,v_2,v_3,v_4) =  \frac{7 v_2}{2560}-\frac{v_1^2}{11520}+\frac{v_4}{1152 v_1^2}-\frac{v_3}{320 v_1}+\frac{v_2^3}{360 v_1^4}+\frac{11 v_2^2}{3840 v_1^2}-\frac{7 v_3 v_2}{1920 v_1^3},
\eeq
etc. 
The algorithm for computing the functions $H_g$ can be found in \cite{DLYZ}. They were used in the construction of the associated integrable hierarchy via the quasi-triviality transformation approach \cite{DZ}.

Let us now proceed to the higher genus terms for the random matrix free energy (recall that only even couplings are allowed).

\begin{theorem} \label{Fgthm}
There exist functions $F_g(v,v_1, \dots, v_{3g-2})$, $g\geq 1$ of independent variables 
$v$, $v_1$, $v_2$, \dots such that
\beq\label{quasiF}
\F_g(x, {\bf s})=F_g \left( u(x, {\bf s}), \frac{\pal u(x, {\bf s})}{\pal x}, \dots, \frac{\pal^{3g-2}u(x, {\bf s})}{\pal x^{3g-2}}\right), \quad g\geq 1.
\eeq
Here
$$
u(x,{\bf s})=\frac{\pal^2 \F_0(x,{\bf s})}{\pal x^2}=\log w(x,{\bf s}).
$$
Recall that the function $w(x,{\bf s})$ is determined from eq.\,\eqref{weven}.
\end{theorem}

Explicitly
\beq\label{F1}
F_1(v, v_1)=\frac1{12} \log v_1 +{\rm const}
\eeq
with const=$\frac{i\pi}{24}+\zeta'(-1)$,
\beq\label{F2}
F_2(v_1, v_2, v_3, v_4)=- \frac{v_2}{480}-\frac{v_1^2}{2880}+\frac{v_4}{288 \, v_1^2} - \frac{v_3}{480 \, v_1} + \frac{v_2^3}{90 \, v_1^4}+\frac{v_2^2}{960 \, v_1^2}-\frac{7 v_3 v_2}{480 \, v_1^3}
\eeq
etc. For any $g\geq 2$ the function $F_g$ is a polynomial in the variables $v_2$, \dots, $v_{3g-2}$ with coefficients in $\mathbb Q\left[ v_1, v_1^{-1}\right]$. 

Using the fact that $\pal_{t_0}=\pal_x$ (see Section \ref{genus12sub} below) along with the standard expansion
$$
\cosh\left(\frac{\epsilon\, \pal_x}2\right)=1+\sum_{n\geq 1} \frac1{(2n)!}\left( \frac{\epsilon}2\right)^{2n} \pal_x^{2n}
$$
we recast
the Main Conjecture for $g\geq 1$ into a sequence of the following relationships 
between the functions $F_g$ and $H_g$ 
\beq\label{FH1}
F_1=2 H_1 +\frac{v}8+{\rm const}
\eeq
and, for $g\geq 2$
\beq\label{FHg}
F_g(v_1,\dots,v_{3g-2}) = \frac{v_{2g-2}}{2^{2g} \, (2g)!}  +\frac{D_0^{2g-2} H_1 (v;v_1)}{2^{2g-3} \, (2g-2)!} + \sum_{m=2}^{g} \frac{2^{3m-2g}}{(2g-2m)!}D_0^{2(g-m)} H_m (v_1,\dots,v_{3m-2}) 
\eeq
where the operator $D_0$ is defined by
$$
D_0=v_1\frac{\pal}{\pal v} +\sum_{k\geq 1} v_{k+1}\frac{\pal}{\pal v_k}.
$$
For example,
\beq\label{FH2}
F_2(v_1,v_2,v_3,v_4) = 4 H_2(v_1,v_2,v_3,v_4) + \frac 14 D_0^2{H_1} + \frac1{384} v_2.
\eeq
Eqs.\,\eqref{FH1}, \eqref{FH2} can be easily verified (see below). In order to verify validity of eqs.\,\eqref{FHg} for any $g\geq 2$ we write a conjectural explicit expression for the functions $F_g(v_1, \dots, v_{3g-2})$ responsible for the genus $g$ random matrix free energies. This will be done in the next subsection.

\setcounter{equation}{0}
\setcounter{theorem}{0}
\subsection{An explicit expression for $F_g$}
We first  recall some notations.  $\mathbb{Y}$ will denote the set of all partitions. For any partition $ \lambda\in\mathbb{Y}$ denote by
$\ell(\lambda)$ the {\em length} of $\lambda$,
by $\lambda_1, \lambda_2,  \dots, \lambda_{\ell(\lambda)}$ 
the non-zero components, $|\lambda|=\lambda_1+\dots+\lambda_{\ell(\lambda)}$ 
the {\em weight},  and by $m_i(\lambda)$ the {\em multiplicity} of $i$ in $\lambda$. Put
$
m(\lambda)!:=\prod_{i\geq 1} m_i(\lambda)!.
$
The set of all partitions of weight $k$ will be denoted by  $\mathbb{Y}_k$. 
For an arbitrary sequence of variables $v_1,v_2,\dots,$ denote $v_\lambda=v_{\lambda_1}\cdots v_{\lambda_{\ell(\lambda)}}.$

\begin{guess} \label{c2} For any
$g\geq 2$, the genus $g$ GUE free energy $F_g$ has the following expression
\eqa
&& \hspace{-6mm}  F_g(v_1,\dots,v_{3g-2}) = \frac{v_{2g-2}}{2^{2g}\, (2g)!}  
+\frac1{2^{2g-3}\, (2g-2)!} D_0^{2g-2}\left( -\frac1{16} v +\frac1{24} \log v_1\right) 
+ \sum_{m=2}^{g} \frac{2^{3m-2g}}{(2g-2m)!}\nn\\
&& \hspace{-6mm} \qquad\qquad\qquad \sum_{k=0}^{3m-3}  \sum_{k_1+k_2+k_3=k\atop 0\leq k_1,k_2,k_3\leq m} \frac{(-1)^{k_2+k_3}}{2^{k_1}}\sum_{\rho,\mu\in\mathbb{Y}_{3m-3-k}} \frac{\langle \lambda_{k_1}\lambda_{k_2}\lambda_{k_3}
\tau_{\rho+1}\rangle_g
}{m(\rho)!} \, Q^{\rho\mu}\, D_0^{2g-2m} 
\left(\frac{v_{\mu+1}}{v_1^{\ell(\mu)+m-1-k}}\right) \nn\\ \label{c2-form}
\eeqa
where for a partition $\mu=(\mu_1,\dots,\mu_{\ell})$,
$\mu+1$ denotes the partition $(\mu_1+1,\dots,\mu_{\ell}+1)$, 
$Q^{\rho\mu}$ is the so-called Q-matrix defined by
\beq
Q^{\rho\mu}=(-1)^{\ell(\rho)}\sum_{\substack{\mu^1\in \mathbb{Y}_{\lambda_1},\dots,\mu^{\ell(\rho)}\in\mathbb{Y}_{\lambda_{\ell(\rho)}} \\ \cup_{q=1}^{\ell(\rho)} \mu^q=\mu}} \prod_{q=1}^{\ell(\rho)}
\frac{\left(\rho_q+\ell(\mu^q)\right)!\,(-1)^{\ell(\mu^q)}}{m(\mu^q)! \prod_{j=1}^\infty (j+1)!^{m_j(\mu^q)}}. \nn
\eeq
In this formula we have used the notation
$$
\langle \lambda_{k_1}\lambda_{k_2}\lambda_{k_3}
\tau_{\nu}\rangle_g:=\int_{\overline{\mathcal{M}}_{g,\ell}}\lambda_{k_1}\lambda_{k_2}\lambda_{k_3} \psi_1^{\nu_1}\dots \psi_{\ell}^{\nu_\ell}, \qquad \forall\, \nu=\left( \nu_1, \dots, \nu_\ell\right)\in \mathbb{Y}.
$$
\end{guess}
Details about $Q$-matrix can be found in \cite{DY2}.
Conj.\,\ref{c2} indicates that the the special cubic Hodge integrals  \eqref{chbcy} naturally appear in the expressions for the higher genus terms of GUE free energy.

\paragraph{Organization of the paper}  
In Sect.\,\ref{DZapproach} we review the approach of \cite{DZ,icmp} to 
the GUE free energy, and prove
Prop.\,\ref{F0thm} and Thm.\,\ref{Fgthm}. 
In Sect.\,\ref{proof012} we verify Conj.\,\ref{c1} and Conj.\,\ref{c2} up to 
the genus $2$ approximation, and give explicit formulae of $\F_g$ for 
$g=3,4,5.$

\paragraph{Acknowledgements}
We wish to thank Si-Qi Liu and Youjin Zhang for helpful discussions.

\section{GUE free energy with even valencies} \label{DZapproach}
\setcounter{equation}{0}
\setcounter{theorem}{0}
\subsection{Calculating the GUE free energy from Frobenius manifold of $\mathbb{P}^1$ topological $\sigma$-model}
It is known that the GUE partition function $Z_N$ (with even and odd couplings) is the tau-function of a particular solution to the Toda lattice hierarchy.
Using this fact, one of the authors in \cite{icmp} developed an efficient algorithm of calculating of GUE free energy, which is an application of the general approach of \cite{DZ,Du1} for the particular example of the two-dimensional Frobenius manifold 
with potential
$$
F=\frac12 u\, v^2 + e^u.
$$
(Warning: only in this section, the notation $v$ is different from that of the Introduction.)
In this section, we give a brief reminder of this approach referring the readers to \cite{DZ,DZ-toda,icmp} for details.

Introduce two analytic functions $\theta_1(u,v;z), \theta_2(u,v;z)$ as follows
\eqa
&& \theta_1(u,v;z) = -2 \, e^{z v} \, \sum_{m=0}^\infty  \left(-\frac12 u +c_m \right) e^{m u} \frac{z^{2m}} {m!^2}=:\sum_{p\geq 0} \theta_{1,p}(u,v) \, z^p \\
&& \theta_2(u,v;z) = z^{-1} \, \left( \sum_{m\geq 0} e^{m u+ z v} \frac{z^{2m}}{(m!)^2} -1 \right)=:\sum_{p\geq 0} \theta_{2,p}(u,v) \, z^p.
\eeqa
Here $c_m=\sum_{k=1}^m\frac1{k}$ denotes the $m$-th harmonic number.

Note that, as in the Introduction, we will only consider the GUE partition function with \textit{even} couplings. 
The corresponding Euler--Lagrange equation \cite{icmp,Du1,DZ} reads
\eqa
&&  x  - w +  \sum_{k\geq 1} (2k)!\, s_k\, \sum_{m=1}^{k} m\, w^m  \, \frac{v^{2k-2m}}{(2k-2m)! \, m!^2}  = 0 \label{elgue1}  \\
&&  -v + \sum_{k\geq 1} (2k)!\, s_k\, \sum_{m=0}^{k-1} w^m \, \frac{v^{2k-1-2m}}{(2k-1-2m)! \, m!^2}  = 0 \label{elgue2}
\eeqa
where $w=e^u$ (as in the Introduction).
Note that we are only interested in the unique series solution $(v(x,{\bf s}),w(x,{\bf s}))$ 
of \eqref{elgue1},\,\eqref{elgue2} such that $v(x,{\bf 0})=0$, $w(x,{\bf 0})=x$.
It is then easy to see from eq.\,\eqref{elgue2} that 
$$v=v(x,{\bf s})\equiv0.$$ 
And eq.\,\eqref{elgue1} becomes
\beq \label{weven-s}
x  - w +  \sum_{m\geq 1}  s_{2m}\,  m\, w^m  \, \frac{(2m)!}{m!^2}  = 0.
\eeq

Define a family of analytic functions $\Omega_{\alpha,p;\beta,q}(u,v)$
by the following generating formula 
\beq\label{two-pt}
\sum_{p,q\geq 0} \Omega_{\alpha,p;\beta,q} z^p y^q= \frac{1}{z+y} \left[\frac{\pal \theta_\alpha(z)}{\pal v} \frac{\pal \theta_\beta(y)}{\pal u} 
+ \frac{\pal \theta_\alpha(z)}{\pal u} \frac{\pal \theta_\beta(y)}{\pal v}  -\delta_{\alpha+\beta,3}\right],\qquad \alpha,\beta=1,2.
\eeq
The genus zero GUE free energy $\mathcal{F}_0(x,{\bf s})$ then has the following expression
\eqa
\mathcal{F}_0 &=& \frac12 \sum_{p,q\geq 2} 
(2p)!(2q)! \, s_{p} s_{q} \, \Omega_{2,2p-1;2,2q-1} 
+ x \sum_{q\geq 1} (2q)! \, s_{q} \, \Omega_{1,0;2,2q-1}  - x \, \Omega_{1,0;2,1}\nn\\
&&  + \frac12 (1-2s_1)^2 \, \Omega_{2,1;2,1}  
+ \sum_{q\geq 2} (2 s_1-1) \, (2q)! \, s_{q} \,  \Omega_{2,1;2,2q-1}
+ \frac12 x^2 \, \Omega_{1,0;1,0}. \label{F0gue}
\eeqa
The higher genus terms in the $1/N$ expansion of the GUE free energy can be determined recursively from the \textit{loop equation} \cite{DZ,icmp} for a sequence of functions
$$
F_g=F_g(u, v, u_1, v_1, \dots, v_{3g-2}, u_{3g-2}), \quad g\geq 1.
$$
This equation has the following form
\eqa
&& \hspace{-6mm} \quad
\sum_{r\geq 0}\left[{\pal \Delta{\cal F}\over \pal v_r} 
\left(v-\lambda\over D\right)_r - 2 {\pal \Delta{\cal F}\over \pal u_r} 
\left(1\over D\right)_r\right] \nn\\
&&\hspace{-6mm} \quad
+\sum_{r\geq 1} \sum_{k=1}^r \left(\begin{array}{c} r \\  k \end{array}\right) 
\left(1\over \sqrt{D}\right)_{k-1} \left[ {\pal \Delta{\cal F}\over \pal v_r}
\left(v-\lambda\over \sqrt{D}\right)_{r-k+1} -2 
{\pal \Delta{\cal F}\over \pal u_r} \left(1\over \sqrt{D}\right)_{r-k+1}\right]
\nn\\
&& \hspace{-6mm}
=D^{-3} e^{u} \left( 4\, e^{u}+(v-\lambda)^2\right)
- \epsilon^2 \sum_{k,l} \left[ \frac14 \, S(\Delta\F,v_k,v_l)
\,  \left(v-\lambda\over \sqrt{D}\right)_{k+1}  \left(v-\lambda\over
\sqrt{D}\right)_{l+1}\right.
\nn\\
&& \hspace{-4mm}
- S(\Delta\F,v_k,u_l) \left(v-\lambda\over
\sqrt{D}\right)_{k+1} \left(1\over \sqrt{D}\right)_{l+1}
  \left. + S(\Delta\F,u_k,u_l) \,
\left(1\over \sqrt{D}\right)_{k+1} \left(1\over \sqrt{D}\right)_{l+1}\right]
\nn\\
&& \hspace{-4mm}
-\frac{\epsilon^2} 2 \sum_k  \left[ {\pal \Delta{\cal F}\over \pal v_k}
  {4 \, e^{u}
(v-\lambda) \, u_1 - T\, v_1 \over D^3}
 +{\pal \Delta{\cal F}\over \pal u_k} 
 {4 \, (v-\lambda) \, v_1 -T\, u_1
\over D^3}\right](e^{u})_{k+1} \label{loop}
\eeqa
where $\triangle \F= \sum_{g\geq 1} \epsilon^{2g} F_g$,  
$D= (v-\lambda)^2 -4\, e^{u}, $
$T= (v-\lambda)^2 + 4\, e^{u},$
 $S(f,a,b):=
{\pal^2 f\over \pal a \pal b}+
{\pal f \over \pal a}
{\pal f \over \pal b}, $ and $f_r$ stands for $\pal_x^r(f).$
Solution $\Delta \F$ of \eqref{loop} exists and is unique up to an additive constant. $F_g$ is a polynomial in $u_2,v_2,\dots,u_{3g-2},v_{3g-2}.$ For $g\geq 2,$ $F_g$ is a rational function 
of $u_1,v_1.$ Then \cite{DZ} the genus $g$ term in the expansion \eqref{genusF}, in the particular case of even couplings only, reads
$$
\mathcal{F}_g(x,{\bf s)} = F_g \left( u(x, {\bf s}), v=0, \frac{\pal u(x, {\bf s})}{\pal x},v_x=0, \dots, \frac{\pal^{3g-2}u(x, {\bf s})}{\pal x^{3g-2}}, v_{3g-2}=0\right)
,\qquad g\geq1
$$

This procedure will be used in the next subsection. 

\setcounter{equation}{0}
\setcounter{theorem}{0}
\subsection{Proof of Prop.\,\ref{F0thm}, Thm.\,\ref{Fgthm}}

\noindent \textit{Proof} of Prop.\,\ref{F0thm}.  Noting that
\eqa
&& \theta_2(u,0,z) = z^{-1} \, \Big( \sum_{m\geq 0} w^m\, \frac{z^{2m}}{(m!)^2} -1 \Big), \nn \\
&& \pal_v \theta_2(u,0,z)  = \sum_{m\geq 0} w^m\, \frac{z^{2m}}{(m!)^2},\qquad  \pal_u \theta_2(u,0,z)  = \sum_{m\geq 0} m w^m\, \frac{z^{2m-1}}{(m!)^2}\nn
\eeqa
and using \eqref{two-pt} we have  
$$
 \sum_{p,q\geq 0} \Omega_{2,p;2,q} z^p y^q = \frac{\sum_{m\geq 0} w^m\, \frac{z^{2m}}{(m!)^2}  \, \sum_{m\geq 0} m w^m\, \frac{y^{2m-1}}{(m!)^2} 
+ \sum_{m\geq 0} m w^m\, \frac{z^{2m-1}}{(m!)^2}  \,   \sum_{m\geq 0} w^m\, \frac{y^{2m}}{(m!)^2}}{z+y}. 
$$
It follows that if $p+q$ is odd then
$
\Omega_{2,p;2,q}
$
vanishes; otherwise, we have
\eqa
\Omega_{2,p;2, q} &=& \frac{w^{\frac{p+q}2+1}}{\left(1+\frac{p+q}2\right) \, \left[\left(\frac p2\right)!\right]^2\, \left[ \left(\frac q2\right)!\right]^2 } ,\qquad p,q \mbox{ are both even}; \\
\Omega_{2,p;2, q} &=& \frac{\frac{p+1}2 \frac{q+1}2 w^{\frac{p+q}2+1}}{\left(1+\frac{p+q}2\right) \, \left[\left(\frac {p+1}2\right)!\right]^2\, \left[ \left(\frac {q+1}2\right)!\right]^2 } ,\qquad p,q \mbox{ are both odd}. 
\eeqa
Substituting these expressions in \eqref{F0gue} we obtain
\eqa
&&\mathcal{F}_0 =\frac12 x^2 u+  \frac12 \sum_{k_1,k_2\geq 0} (2k_1+2)!(2k_2+2)!s_{k_1+1} s_{k_2+1} \frac{(k_1+1)(k_2+1) \,  w^{k_1+k_2+2} } {(k_1+k_2+2) \, \left[\left(k_1+1\right)!\right]^2\, \left[ \left(k_2+1\right)!\right]^2 }   \nn \\ 
&& + x \sum_{k\geq 0} (2k+2)! s_{k+1} \frac{w^{k+1}}{(k+1)!^2} - x \, w  + (1-4s_1) \frac{w^2}4 
- \sum_{k\geq 1}(2k+2)! s_{k+1}  \frac{(k+1) \, w^{k+2}}{(k+2) \left[ \left(k+1\right)!\right]^2 }. \nn
\eeqa
Equation \eqref{weven} is already proved in \eqref{weven-s}.
The proposition is proved.  \epf

\noindent \textit{Proof} of Theorem \ref{Fgthm}.
For $g=1,2$, taking $v=v_1=v_2=\dots=0$ in the general expressions of 
$F_g(u,v,u_1,v_1,\dots,u_{3g-2},v_{3g-2})$  
\cite{DZ,DZ-toda} one obtains \eqref{F1} and \eqref{F2}. For any $g\geq 1$,  
the existence of $F_g(u,u_1,\dots,u_{3g-2})$  such that 
$$
\F_g(x, {\bf s})=F_g \left( u(x, {\bf s}), \frac{\pal u(x, {\bf s})}{\pal x}, \dots, \frac{\pal^{3g-2}u(x, {\bf s})}{\pal x^{3g-2}}\right)
$$
is a direct result of \cite{DZ,DZ-toda} when taking $v=v_1=v_2=\dots=0$ in 
$F_g(u,v,u_1,v_1,\dots,u_{3g-2},v_{3g-2})$.
\epf

\section{Verification of the Main Conjecture for low genera} \label{proof012}
\setcounter{equation}{0}
\setcounter{theorem}{0}
\subsection{Genus 0}
Recall that the genus zero cubic Hodge free energy can be expressed as
$$
\mathcal{H}_0({\bf t}) = \frac12  \sum_{i,j\geq 0}  \tilde t_i \, \tilde t_j \, \Omega_{i;j}(v({\bf t})).
$$
where $\tilde t_i= t_i -\delta_{i,1},$ ${\bf t}=(t_0,t_1,t_2,\dots)$, $\Omega_{i;j}$ are polynomials in $v$ given by
 $$\Omega_{i;j}(v)=\frac{v^{i+j+1}}{(i+j+1) \, i! \, j!},$$
 and $v({\bf t})$ is the unique series solution to the following Euler--Lagrange equation of the one-dimensional Frobenius manifold
 $$
 v = \sum_{i\geq 0} t_i \frac{v^i}{i!}.
 $$
(Warning: the above $v$ is the flat coordinate of the one-dimensional Frobenius manifold; avoid confusing with 
$v$ in Section 2 where $(u,v)$ are flat coordinates of the two-dimensional Frobenius manifold of $\mathbb{P}^1$ topological 
$\sigma$-model.)

Let us consider the following substitution of time variables
$$
t_i = \sum_{k\geq 1} k^{i+1} \bar{s}_k  - 1 + \delta_{i,1} + x \cdot \delta_{i,0},\qquad i\geq 0.
$$
Note that with this substitution the cubic Hodge free energies will be considered to be expanded at $x=1$. We have $\tilde t_i = \sum_{k\geq 1} k^{i+1} \bar{s}_k - 1 + x \cdot \delta_{i,0},$ and so
\eqa
\mathcal{H}_0 & = & \frac12  \sum_{i,j\geq 0}  \tilde t_i \, \tilde t_j \, \Omega_{i;j}(v({\bf t})) \nn \\
&=&  \frac12  \sum_{i,j\geq 0}  \left(\sum_{k_1\geq 1} k_1^{i+1} \bar{s}_{k_1} - 1  + x \cdot \delta_{i,0}\right) \,\left(\sum_{k_2\geq 1} k_2^{j+1} \bar{s}_{k_2} - 1 + x \cdot \delta_{j,0} \right)\,\frac{v^{i+j+1}}{(i+j+1) \, i! \, j!} \nn\\
&=& \frac12  \sum_{i,j\geq 0}  \sum_{k_1,k_2\geq 1} k_1^{i+1} \, k_2^{j+1} \,  \bar{s}_{k_1}  \, \bar{s}_{k_2} \,\frac{v^{i+j+1}}{(i+j+1) \, i! \, j!} - 
 \sum_{i,j\geq 0}  \sum_{k_1\geq 1} k_1^{i+1}  \,  \bar{s}_{k_1}  \,\frac{v^{i+j+1}}{(i+j+1) \, i! \, j!} \nn\\
 && +  x \, \sum_{i,j\geq 0}  \sum_{k_1\geq 1} k_1^{i+1}  \,  \bar{s}_{k_1} \, \delta_{j,0}  \,\frac{v^{i+j+1}}{(i+j+1) \, i! \, j!} +\frac12 \sum_{i,j\geq 0}\frac{v^{i+j+1}}{(i+j+1) \, i! \, j!}\nn\\
 && - x \, \sum_{i,j\geq 0}  \, \delta_{j,0}  \,\frac{v^{i+j+1}}{(i+j+1) \, i! \, j!}  + \frac{x^2} 2\sum_{i,j\geq 0} \delta_{i,0}  \, \delta_{j,0}  \,\frac{v^{i+j+1}}{(i+j+1) \, i! \, j!} .
\eeqa
We simplify it term by term:
\eqa
&& \frac{x^2} 2\sum_{i,j\geq 0} \delta_{i,0}  \, \delta_{j,0}  \,\frac{v^{i+j+1}}{(i+j+1) \, i! \, j!} = \frac{x^2}2 v,\nn\\
&& x \, \sum_{i,j\geq 0}  \, \delta_{j,0}  \,\frac{v^{i+j+1}}{(i+j+1) \, i! \, j!}  = x\, ( e^v -1),\nn\\ 
&& \frac12 \sum_{i,j\geq 0}\frac{v^{i+j+1}}{(i+j+1) \, i! \, j!} = \frac12 \sum_{\ell\geq 0} \sum_{i= 0}^\ell \frac{v^{\ell+1} \ell!}{(\ell+1)! \, i! \, (\ell-i)!} = \frac12 \sum_{\ell\geq 0}
 \frac{v^{\ell+1} \, 2^\ell}{(\ell+1)!} = \frac14 (e^{2v}-1),  \nn \\
&& x \, \sum_{i,j\geq 0}  \sum_{k_1\geq 1} k_1^{i+1}  \,  \bar{s}_{k_1} \, \delta_{j,0}  \,\frac{v^{i+j+1}}{(i+j+1) \, i! \, j!}=x \, \sum_{i\geq 0}  \sum_{k_1\geq 1} k_1^{i+1}  \,  \bar{s}_{k_1}  \,\frac{v^{i+1}}{(i+1)!}=x \,  \sum_{k\geq 1} \,  \bar{s}_{k} \,  (e^{k v}-1),  \nn \\
&& \sum_{i,j\geq 0}  \sum_{k_1\geq 1} k_1^{i+1}  \,  \bar{s}_{k_1}  \,\frac{v^{i+j+1}}{(i+j+1) \, i! \, j!} =  
\sum_{k\geq 1} k\, \bar{s}_{k}  \sum_{\ell\geq 0}  (1+k)^{\ell}   \,\frac{v^{\ell+1}}{(\ell+1)!} =
\sum_{k\geq 1} \frac{k}{1+k}\, \bar{s}_{k} \, \left(e^{(1+k)v}-1\right), \nn\\
&& \frac12  \sum_{i,j\geq 0}  \sum_{k_1,k_2\geq 1} k_1^{i+1} \, k_2^{j+1} \,  \bar{s}_{k_1}  \, \bar{s}_{k_2} \,\frac{v^{i+j+1}}{(i+j+1) \, i! \, j!} = 
\frac12  \sum_{k_1,k_2\geq 1} k_1\,k_2 \,  \bar{s}_{k_1}  \, \bar{s}_{k_2}   \sum_{\ell\geq 0}  
\,\frac{v^{\ell+1}}{(\ell+1)!} (k_1+k_2)^\ell \nn\\
&&\qquad\qquad\qquad\qquad=\frac12  \sum_{k_1,k_2\geq 1} \frac{k_1\,k_2}{k_1+k_2}  \,  \bar{s}_{k_1}  \, \bar{s}_{k_2}  \left(e^{(k_1+k_2)v}-1\right).\nn
\eeqa
Let $w=e^v$. We have
\eqa
\mathcal{H}_0 &= &\frac12  \sum_{k_1,k_2\geq 1} \frac{k_1\,k_2}{k_1+k_2}  \,  \bar{s}_{k_1}  \, \bar{s}_{k_2}  \left(w^{k_1+k_2}-1\right) -\sum_{k\geq 1} \frac{k}{1+k}\, \bar{s}_{k} \, \left(w^{1+k}-1\right) +x \,  \sum_{k\geq 1} \,  \bar{s}_{k} \,  (w^{k}-1) \nn\\
&& + \frac14 (w^{2}-1) -x\, (w -1) + \frac{x^2}2 \log w. \nn
\eeqa

On the other hand, recall from Prop.\,\ref{F0thm} that the genus zero GUE free energy with even couplings has the form
$$
\mathcal{F}_0 = \frac{w^2}4 - x \, w + \sum_{k\geq 1} \bar{s}_k \left(x \, w^k-\frac{k}{k+1} w^{k+1}\right) +\frac12 \sum_{k_1,k_2\geq 1} \frac{k_1k_2}{k_1+k_2} 
\bar{s}_{k_1} \bar{s}_{k_2} w^{k_1+k_2} + \frac{x^2} 2 \log w.
$$
Here $w$ is the power series solution to 
$$
w= x+ \sum_{k\geq 1} \, k\, \bar s_k  \, w^k.
$$
Recall that $w=e^u$; so 
$$
e^u= x+ \sum_{k\geq 1} \, k\, \bar s_k  \, e^{ku}.
$$
Namely,
$$
1 + \sum_{j\geq 1} \frac{u^j}{j!} = x + \sum_{k\geq 1} \, k\, \bar s_k  \, \left(1 + \sum_{j\geq 1} \frac{k^j \, u^j}{j!}\right) .
$$
It follows that 
\beq \label{uveq}
u(x,{\bf s})=v(\bt(x,{\bf s})).
\eeq
We conclude that
\beq
\mathcal{H}_0(\bt(x,{\bf s})) -\mathcal{F}_0(x,{\bf s}) =  -
\frac12  \sum_{k_1,k_2\geq 1} \frac{k_1\,k_2}{k_1+k_2}  \,  \bar{s}_{k_1}  \, \bar{s}_{k_2} + \sum_{k\geq 1} \frac{k}{1+k}\, \bar{s}_{k} -
x \,  \sum_{k\geq 1} \,  \bar{s}_{k}   - \frac14  + x. 
\eeq
This finishes the proof of the genus zero part of the Main Conjecture. 

\setcounter{equation}{0}
\setcounter{theorem}{0}
\subsection{Genus $1,2$} \label{genus12sub}
Note that the substitution \eqref{time-sub}  $$(t_0,t_1,t_2,\dots)\mapsto (x,\bar s_1, \bar s_2,\dots)$$ 
satisfies that
\eqa
&& \frac{\partial }{\partial x}=\frac{\partial}{\partial t_0},\\
&& \frac{\partial }{\partial \bar s_k} = \sum_{i\geq 0} k^{i+1} \frac{\partial }{\partial t_i}, \qquad k\geq 1.
\eeqa
In particular, we have
$$
\frac{\partial v}{\partial t_0}(\bt(x,{\bf s}))= \frac{\partial v(\bt(x,{\bf s}))}{\partial x} = \frac{\pal u(x,{\bf s})}{\pal x}.
$$
The last equality is due to \eqref{uveq}.

Recall, from the algorithm of \cite{DLYZ}, that the genus $1$ special cubic Hodge free energy is given by
$$
H_1(v;v_1) =   \frac1{24} \log v_1  - \frac v{16}.
$$
So 
$$
2H_1(v;v_1) + \frac v8 +\frac{i \pi}{24}+\zeta'(-1) = \frac1{12} \log v_1   +\frac{i \pi}{24}+\zeta'(-1).
$$
This proves the genus $1$ part of the Main Conjecture. 

The genus $2$ term of the special cubic Hodge free energy is given by
$$
H_2(v_1,v_2,v_3,v_4) =  \frac{7 \, v_2}{2560}-\frac{v_1^2}{11520}+\frac{v_4}{1152 \, v_1^2}-\frac{v_3}{320 \, v_1}+\frac{v_2^3}{360 \, v_1^4}+\frac{11 \, v_2^2}{3840 \, v_1^2}-\frac{7 \, v_3 v_2}{1920 \, v_1^3}.
$$
So 
\eqa
4 H_2+ \frac 14 \, D_0^2 \, H_1 + \frac1{384} v_2 &=& 4 H_2(v_1,v_2,v_3,v_4)-\frac5{384} v_2 +\frac1{96} \left[ \frac{v_3}{v_1}-\left(\frac{v_2}{v_1}\right)^2\right] \nn\\
&=& - \frac{v_2}{480}-\frac{v_1^2}{2880}+\frac{v_4}{288 \, v_1^2} - \frac{v_3}{480 \, v_1} + \frac{v_2^3}{90 \, v_1^4}+\frac{v_2^2}{960 \, v_1^2}-\frac{7 v_3 v_2}{480 \, v_1^3}\nn\\
&=& F_2(v_1,v_2,v_3,v_4).\nn
\eeqa
This proves the genus $2$ part of the Main Conjecture. 

\setcounter{equation}{0}
\setcounter{theorem}{0}
\subsection{Genus $3,4$}
Using the Main Conjecture along with the algorithm of \cite{DLYZ}, we obtain the following two statements. 
\begin{guess} The genus $3$ GUE free energy is given by
\eqa
&&\hspace{-6mm} ~ F_3(u_1,\dots,u_7)\nn\\
&&\hspace{-10mm} = \frac{13 u_4}{120960}+\frac{u_2^2}{24192}-\frac{u_1^4}{725760}+\frac{u_7}{10368 u_1^3}-\frac{u_6}{5760 u_1^2}-\frac{u_5}{13440 u_1} \nn\\
&&\hspace{-6mm} -\frac{103 u_4^2}{60480 u_1^4}+\frac{59 u_3^3}{8064 u_1^5}+\frac{u_3^2}{2688 u_1^2}+\frac{u_3 u_1}{12096}-\frac{5 u_2^6}{81 u_1^8}-\frac{13 u_2^5}{1890 u_1^6}\nn\\
&&\hspace{-6mm} +\frac{5 u_2^4}{5376 u_1^4}-\frac{u_2^3}{9072 u_1^2}-\frac{7 u_6 u_2}{5760 u_1^4}-\frac{53 u_3 u_5}{20160 u_1^4}+\frac{353 u_5 u_2^2}{40320 u_1^5}+\frac{u_5 u_2}{840 u_1^3}\nn\\
&&\hspace{-6mm} +\frac{89 u_3 u_4}{40320 u_1^3}-\frac{83 u_4 u_2^3}{1890 u_1^6}-\frac{211 u_4 u_2^2}{40320 u_1^4}+\frac{u_4 u_2}{2016 u_1^2}+\frac{59 u_3 u_2^4}{378 u_1^7}\nn\\
&&\hspace{-6mm} +\frac{1993 u_3 u_2^3}{120960 u_1^5}-\frac{u_3 u_2^2}{576 u_1^3}-\frac{83 u_3^2 u_2^2}{896 u_1^6}+\frac{19 u_3 u_2}{120960 u_1}-\frac{17 u_3^2 u_2}{2240 u_1^4}+\frac{1273 u_3 u_4 u_2}{40320 u_1^5}.\nn\\
&&\hspace{-6mm} \label{f3even}
\eeqa
\end{guess}

\begin{guess} The genus $4$ GUE free energy is given by
\eqa
&&~F_4(u_1,\dots,u_{10}) \nn\\
&&\hspace{-4mm}= \frac{1852 u_2^9}{1215 u_1^{12}}+\frac{151 u_2^8}{675 u_1^{10}}-\frac{101 u_2^7}{12600 u_1^8}-\frac{772 u_3 u_2^7}{135 u_1^{11}}\nn\\
&&+\frac{9904 u_4 u_2^6}{6075 u_1^{10}}-\frac{1165 u_2^6}{1161216 u_1^6}-\frac{2851 u_3 u_2^6}{3600 u_1^9}+\frac{14903 u_3^2 u_2^5}{2160 u_1^{10}}+\frac{70261 u_3 u_2^5}{3225600 u_1^7}\nn\\
&&+\frac{2573 u_4 u_2^5}{10800 u_1^8}+\frac{u_2^5}{7200 u_1^4}-\frac{2243 u_5 u_2^5}{6480 u_1^9}+\frac{195677 u_3^2 u_2^4}{230400 u_1^8}+\frac{3197 u_3 u_2^4}{967680 u_1^5}\nn\\
&&+\frac{12907 u_6 u_2^4}{226800 u_1^8}-\frac{10259 u_4 u_2^4}{1935360 u_1^6}-\frac{22153 u_5 u_2^4}{414720 u_1^7}-\frac{101503 u_3 u_4 u_2^4}{32400 u_1^9}+\frac{1823 u_4^2 u_2^3}{5670 u_1^8}\nn\\
&&+\frac{415273 u_3 u_5 u_2^3}{829440 u_1^8}+\frac{97 u_5 u_2^3}{120960 u_1^5}+\frac{26879 u_6 u_2^3}{2903040 u_1^6}+\frac{u_2^3}{7257600}-\frac{49 u_3 u_2^3}{138240 u_1^3}\nn\\
&&-\frac{5137 u_4 u_2^3}{4354560 u_1^4}-\frac{877 u_3^2 u_2^3}{57600 u_1^6}-\frac{812729 u_3 u_4 u_2^3}{2073600 u_1^7}-\frac{212267 u_7 u_2^3}{29030400 u_1^7}-\frac{305129 u_3^3 u_2^3}{103680 u_1^9}\nn\\
&&+\frac{u_1^2 u_2^2}{460800}+\frac{1379 u_4^2 u_2^2}{34560 u_1^6}+\frac{13138507 u_3^2 u_4 u_2^2}{9676800 u_1^8}+\frac{2417 u_3 u_4 u_2^2}{537600 u_1^5}+\frac{17 u_4 u_2^2}{138240 u_1^2}\nn\\
&&+\frac{2143 u_3 u_5 u_2^2}{34560 u_1^6}+\frac{449 u_5 u_2^2}{1451520 u_1^3}+\frac{2323 u_8 u_2^2}{3225600 u_1^6}-\frac{2623 u_3^2 u_2^2}{967680 u_1^4}-\frac{443 u_6 u_2^2}{9676800 u_1^4}\nn\\
&&-\frac{667 u_7 u_2^2}{537600 u_1^5}-\frac{192983 u_3^3 u_2^2}{691200 u_1^7}-\frac{60941 u_3 u_6 u_2^2}{1075200 u_1^7}-\frac{171343 u_4 u_5 u_2^2}{1935360 u_1^7}+\frac{22809 u_3^4 u_2}{71680 u_1^8}\nn\\
&&+\frac{1747 u_3^3 u_2}{806400 u_1^5}+\frac{7 u_3^2 u_2}{38400 u_1^2}+\frac{9221 u_5^2 u_2}{1935360 u_1^6}+\frac{17 u_1 u_3 u_2}{3225600}+\frac{78533 u_3^2 u_4 u_2}{691200 u_1^6}\nn\\
&&+\frac{18713 u_3 u_4 u_2}{14515200 u_1^3}+\frac{15179 u_4 u_6 u_2}{1935360 u_1^6}+\frac{20639 u_3 u_7 u_2}{4838400 u_1^6}+\frac{37 u_8 u_2}{302400 u_1^4}-\frac{u_4 u_2}{86400}\nn\\
&&-\frac{11 u_5 u_2}{362880 u_1}-\frac{923 u_6 u_2}{14515200 u_1^2}-\frac{113 u_7 u_2}{9676800 u_1^3}-\frac{55 u_4^2 u_2}{387072 u_1^4}-\frac{419 u_3 u_5 u_2}{1935360 u_1^4}\nn\\
&&-\frac{1411 u_4 u_5 u_2}{138240 u_1^5}-\frac{7 u_9 u_2}{138240 u_1^5}-\frac{1751 u_3 u_6 u_2}{268800 u_1^5}-\frac{12035 u_3^2 u_5 u_2}{96768 u_1^7}-\frac{44201 u_3 u_4^2 u_2}{276480 u_1^7}\nn\\
&&+\frac{1549 u_3^4}{115200 u_1^6}+\frac{937 u_3^3}{2903040 u_1^3}+\frac{229 u_4^3}{62208 u_1^6}+\frac{19 u_5^2}{46080 u_1^4}+\frac{u_1^3 u_3}{691200}+\frac{949 u_3 u_4 u_5}{55296 u_1^6}\nn\\
&&+\frac{59 u_3^2 u_6}{10752 u_1^6}+\frac{73 u_4 u_6}{107520 u_1^4}+\frac{1777 u_3 u_7}{4838400 u_1^4}+\frac{143 u_7}{14515200 u_1}+\frac{31 u_8}{9676800 u_1^2}+\frac{u_{10}}{497664 u_1^4}\nn\\
&&-\frac{u_3^2}{115200}-\frac{u_1 u_5}{138240} -\frac{73 u_6}{29030400}-\frac{u_1^6}{43545600}-\frac{19 u_1^2 u_4}{87091200}-\frac{137 u_3 u_4}{2073600 u_1} \nn\\
&&-\frac{239 u_3 u_5}{1451520 u_1^2}-\frac{661 u_4^2}{5806080 u_1^2}-\frac{u_9}{138240 u_1^3}-\frac{17 u_4 u_5}{387072 u_1^3}-\frac{89 u_3 u_6}{3225600 u_1^3}-\frac{709 u_3^2 u_4}{3225600 u_1^4} \nn\\
&&-\frac{1291 u_3 u_4^2}{138240 u_1^5}-\frac{1001 u_3^2 u_5}{138240 u_1^5}-\frac{197 u_5 u_6}{387072 u_1^5}-\frac{163 u_3 u_8}{967680 u_1^5}-\frac{2069 u_4 u_7}{5806080 u_1^5}-\frac{2153 u_3^3 u_4}{28800 u_1^7}.\nn \\
&& \label{f4even} 
\eeqa
\end{guess}

We also computed the genus $5$ free energy; see in Appendix \ref{appa}. 

For the particular examples of enumerating squares, hexagons, octagons on a genus $g$ surface,
 one can use 
\eqref{f3even},\,\eqref{f4even},\,\eqref{F5} to obtain the combinatorial numbers. We checked that  
these numbers agree with those in \cite{DY1}. This gives some evidences of validity of
 the Main Conjecture for $g=3,4,5$.

\begin{remark} 
The genus $1,2,3$ terms of the GUE free energy with even couplings were also derived in 
\cite{E,EMP,Pierce} for the particular case of only one nonzero coupling (i.e., in the framework of
enumeration of $2m$-gons).  To the best of our knowledge, explicit 
formulae for higher genus ($g\geq 4$) terms, even
in the case of the particular examples, were not available in the literature.
\end{remark}

\appendix
\section{Explicit formula for $F_5$} \label{appa}
\eqa
&& \hspace{-6mm} F_5(u_1,\dots,u_{13})= \nn\\
&&  -\frac{109514 u_2^{12}}{1215 u_1^{16}}-\frac{1352 u_2^{11}}{81 u_1^{14}}+\frac{181628 u_3 u_2^{10}}{405 u_1^{15}}-\frac{2593 u_2^{10}}{9450 u_1^{12}} 
 +\frac{42691 u_3 u_2^9}{540 u_1^{13}} \nn\\
 && +\frac{16091 u_2^9}{187110 u_1^{10}}-\frac{93460 u_4 u_2^9}{729 u_1^{14}}+\frac{600763 u_3 u_2^8}{415800 u_1^{11}}+\frac{134599 u_5 u_2^8}{4860 u_1^{13}}-\frac{274289 u_2^8}{255467520 u_1^8} \nn\\
 && -\frac{567199 u_4 u_2^8}{24300 u_1^{12}}-\frac{1322159 u_3^2 u_2^8}{1620 u_1^{14}}+\frac{391519 u_3 u_4 u_2^7}{972 u_1^{13}}+\frac{2471441 u_5 u_2^7}{475200 u_1^{11}}-\frac{151 u_2^7}{399168 u_1^6}\nn\\
 && -\frac{999473 u_3 u_2^7}{2838528 u_1^9}-\frac{2825 u_4 u_2^7}{5544 u_1^{10}}-\frac{1149739 u_3^2 u_2^7}{8640 u_1^{12}}-\frac{161353 u_6 u_2^7}{34020 u_1^{12}}+\frac{12916717 u_3^3 u_2^6}{19440 u_1^{13}}\nn\\
 && +\frac{319877 u_3 u_2^6}{159667200 u_1^7}+\frac{31781177 u_3 u_4 u_2^6}{475200 u_1^{11}}+\frac{4549471 u_4 u_2^6}{42577920 u_1^8}+\frac{515032871 u_5 u_2^6}{3832012800 u_1^9}\nn\\
 && +\frac{8477461 u_7 u_2^6}{12830400 u_1^{11}}+\frac{263 u_2^6}{23950080 u_1^4}-\frac{27484783 u_6 u_2^6}{29937600 u_1^{10}}-\frac{1098923921 u_3^2 u_2^6}{425779200 u_1^{10}} \nn\\
 && -\frac{5713573 u_3 u_5 u_2^6}{77760 u_1^{12}}-\frac{12051881 u_4^2 u_2^6}{255150 u_1^{12}}+\frac{173831501 u_3^3 u_2^5}{1824768 u_1^{11}}+\frac{6013615 u_3^2 u_2^5}{12773376 u_1^8}\nn\\
 && +\frac{28957 u_3 u_2^5}{21288960 u_1^5}+\frac{276125491 u_3 u_4 u_2^5}{182476800 u_1^9}+\frac{10243 u_4 u_2^5}{239500800 u_1^6}+\frac{53761259 u_4 u_5 u_2^5}{3326400 u_1^{11}}\nn\\
 && +\frac{102985067 u_3 u_6 u_2^5}{9979200 u_1^{11}}+\frac{84091529 u_7 u_2^5}{638668800 u_1^9}+\frac{u_2^5}{633600 u_1^2}-\frac{11532541 u_5 u_2^5}{479001600 u_1^7}-\frac{5117003 u_6 u_2^5}{182476800 u_1^8}\nn\\
 &&-\frac{19847231 u_4^2 u_2^5}{2494800 u_1^{10}}-\frac{2268769 u_8 u_2^5}{29937600 u_1^{10}}-\frac{1125727817 u_3 u_5 u_2^5}{91238400 u_1^{10}}-\frac{14531719 u_3^2 u_4 u_2^5}{36288 u_1^{12}}\nn\\
 &&+\frac{2322129119 u_3^3 u_2^4}{1277337600 u_1^9}+\frac{1211 u_3^2 u_2^4}{2027520 u_1^6}+\frac{87822499 u_3 u_4^2 u_2^4}{1197504 u_1^{11}}+\frac{728671781 u_3^2 u_5 u_2^4}{12773376 u_1^{11}}\nn\\
 &&+\frac{1046529431 u_4 u_5 u_2^4}{383201280 u_1^9}+\frac{1108533611 u_3 u_6 u_2^4}{638668800 u_1^9}+\frac{231617 u_6 u_2^4}{54743040 u_1^6}+\frac{1215667 u_7 u_2^4}{255467520 u_1^7}+\frac{1096859 u_9 u_2^4}{153280512 u_1^9}\nn\\
 &&-\frac{29 u_2^4}{127733760}-\frac{47 u_3 u_2^4}{1596672 u_1^3}-\frac{29783 u_4 u_2^4}{63866880 u_1^4}-\frac{24971 u_5 u_2^4}{127733760 u_1^5}-\frac{443511251 u_3 u_4 u_2^4}{1916006400 u_1^7}\nn\\
 &&-\frac{68301953 u_3 u_5 u_2^4}{212889600 u_1^8}-\frac{5935327 u_8 u_2^4}{383201280 u_1^8}-\frac{133150489 u_4^2 u_2^4}{638668800 u_1^8}-\frac{5682077 u_4 u_6 u_2^4}{2721600 u_1^{10}}\nn\\
 &&-\frac{1095679399 u_3^2 u_4 u_2^4}{19353600 u_1^{10}}-\frac{154991051 u_3 u_7 u_2^4}{136857600 u_1^{10}}-\frac{4872743377 u_5^2 u_2^4}{3832012800 u_1^{10}}-\frac{547560589 u_3^4 u_2^4}{2322432 u_1^{12}}\nn\\
 &&+\frac{u_1^2 u_2^3}{3991680}+\frac{69761129 u_3 u_4^2 u_2^3}{6967296 u_1^9}+\frac{75292039 u_4^2 u_2^3}{2874009600 u_1^6}+\frac{8817996169 u_3^3 u_4 u_2^3}{63866880 u_1^{11}}\nn\\
 &&+\frac{1049 u_4 u_2^3}{119750400 u_1^2}+\frac{2479931059 u_3^2 u_5 u_2^3}{319334400 u_1^9}+\frac{38370113 u_3 u_5 u_2^3}{958003200 u_1^6}+\frac{51887531 u_4 u_5 u_2^3}{638668800 u_1^7}\nn\\
 &&+\frac{2279 u_5 u_2^3}{19160064 u_1^3}+\frac{16316161 u_3 u_6 u_2^3}{319334400 u_1^7}+\frac{25527371 u_5 u_6 u_2^3}{87091200 u_1^9}+\frac{6169 u_6 u_2^3}{76640256 u_1^4}+\frac{791123 u_4 u_7 u_2^3}{3870720 u_1^9}\nn\\
 &&+\frac{369691943 u_3 u_8 u_2^3}{3832012800 u_1^9}+\frac{284539 u_9 u_2^3}{191600640 u_1^7}-\frac{767 u_3 u_2^3}{212889600 u_1}-\frac{2141 u_3^2 u_2^3}{1520640 u_1^4}-\frac{42149 u_7 u_2^3}{71850240 u_1^5} \nn\\
 && -\frac{1229917 u_3 u_4 u_2^3}{638668800 u_1^5}-\frac{52781 u_8 u_2^3}{79833600 u_1^6}-\frac{10864759 u_3^3 u_2^3}{47900160 u_1^7}-\frac{2883761 u_4 u_6 u_2^3}{8294400 u_1^8} \nn\\
 && -\frac{764936639 u_3^2 u_4 u_2^3}{638668800 u_1^8}-\frac{179669059 u_3 u_7 u_2^3}{958003200 u_1^8}-\frac{573019 u_{10} u_2^3}{1045094400 u_1^8}-\frac{406451147 u_5^2 u_2^3}{1916006400 u_1^8} \nn\\
 &&-\frac{1848263 u_4^3 u_2^3}{467775 u_1^{10}}-\frac{626921467 u_3^2 u_6 u_2^3}{106444800 u_1^{10}}-\frac{3430962641 u_3^4 u_2^3}{127733760 u_1^{10}}-\frac{35338084651 u_3 u_4 u_5 u_2^3}{1916006400 u_1^{10}}\nn\\
 &&+\frac{419193 u_3^5 u_2^2}{14336 u_1^{11}}+\frac{u_1^4 u_2^2}{6386688}+\frac{23669 u_3^2 u_2^2}{1277337600 u_1^2}+\frac{1783923 u_3 u_4^2 u_2^2}{7884800 u_1^7}+\frac{50047 u_4^2 u_2^2}{106444800 u_1^4}\nn\\
 &&+\frac{1238492531 u_3 u_5^2 u_2^2}{1277337600 u_1^9}+\frac{181880015 u_3^3 u_4 u_2^2}{12773376 u_1^9}+\frac{73978651 u_3^2 u_4 u_2^2}{638668800 u_1^6}+\frac{3337 u_3 u_4 u_2^2}{4561920 u_1^3}\nn\\
 &&+\frac{5302619 u_3^2 u_5 u_2^2}{30412800 u_1^7}+\frac{1591687897 u_4^2 u_5 u_2^2}{1277337600 u_1^9}+\frac{30953 u_3 u_5 u_2^2}{45619200 u_1^4}+\frac{203383903 u_3 u_4 u_6 u_2^2}{127733760 u_1^9}\nn\\
 &&+\frac{29758507 u_5 u_6 u_2^2}{638668800 u_1^7}+\frac{275874283 u_3^2 u_7 u_2^2}{638668800 u_1^9}+\frac{2069131 u_4 u_7 u_2^2}{63866880 u_1^7}+\frac{289 u_3 u_8 u_2^2}{19008 u_1^7}+\frac{60127 u_8 u_2^2}{958003200 u_1^4}\nn\\
 &&+\frac{3373 u_9 u_2^2}{45619200 u_1^5}+\frac{11437 u_{11} u_2^2}{348364800 u_1^7}-\frac{5 u_1 u_3 u_2^2}{2128896}-\frac{19 u_4 u_2^2}{54743040}-\frac{7753 u_5 u_2^2}{3832012800 u_1}-\frac{223 u_6 u_2^2}{9123840 u_1^2}\nn\\
 &&-\frac{247 u_7 u_2^2}{11612160 u_1^3}-\frac{317077 u_3 u_6 u_2^2}{63866880 u_1^5}-\frac{998201 u_3^3 u_2^2}{638668800 u_1^5}-\frac{5091301 u_4 u_5 u_2^2}{638668800 u_1^5}-\frac{742733 u_5^2 u_2^2}{106444800 u_1^6}\nn\\
 &&-\frac{19631 u_{10} u_2^2}{174182400 u_1^6}-\frac{1295627 u_3 u_7 u_2^2}{212889600 u_1^6}-\frac{7277719 u_4 u_6 u_2^2}{638668800 u_1^6}-\frac{20664649 u_3 u_4 u_5 u_2^2}{8870400 u_1^8}-\frac{9854357 u_3^2 u_6 u_2^2}{13305600 u_1^8}\nn\\
 && -\frac{13316641 u_4^3 u_2^2}{26611200 u_1^8}-\frac{1293697 u_5 u_7 u_2^2}{53222400 u_1^8}-\frac{328171 u_3 u_9 u_2^2}{53222400 u_1^8}-\frac{45787691 u_3^4 u_2^2}{106444800 u_1^8}-\frac{6093371 u_6^2 u_2^2}{425779200 u_1^8}\nn\\
 &&-\frac{18837199 u_4 u_8 u_2^2}{1277337600 u_1^8}-\frac{2912894597 u_3^2 u_4^2 u_2^2}{116121600 u_1^{10}}-\frac{1661611993 u_3^3 u_5 u_2^2}{127733760 u_1^{10}}\nn\\
 &&+\frac{14367497 u_3^5 u_2}{7096320 u_1^9}+\frac{3625841 u_3^4 u_2}{127733760 u_1^6}+\frac{4721 u_3^3 u_2}{12773376 u_1^3}+\frac{274614007 u_3 u_4^3 u_2}{182476800 u_1^9}+\frac{7368997 u_3 u_5^2 u_2}{70963200 u_1^7}\nn\\
 &&+\frac{927517 u_5^2 u_2}{1916006400 u_1^4}+\frac{17 u_1^3 u_3 u_2}{53222400}+\frac{2872733 u_3^3 u_4 u_2}{13305600 u_1^7}+\frac{20527 u_3^2 u_4 u_2}{14192640 u_1^4}+\frac{85455011 u_4^2 u_5 u_2}{638668800 u_1^7}\nn\\
 &&+\frac{1121508611 u_3^2 u_4 u_5 u_2}{319334400 u_1^9}+\frac{159183659 u_3^3 u_6 u_2}{212889600 u_1^9}+\frac{15535133 u_3 u_4 u_6 u_2}{91238400 u_1^7}+\frac{1516703 u_4 u_6 u_2}{1916006400 u_1^4}\nn\\
 &&+\frac{2867 u_5 u_6 u_2}{1814400 u_1^5}+\frac{5833 u_6 u_2}{3832012800}+\frac{916913 u_3^2 u_7 u_2}{19958400 u_1^7}+\frac{49661 u_3 u_7 u_2}{119750400 u_1^4}+\frac{698809 u_4 u_7 u_2}{638668800 u_1^5}\nn\\
 &&+\frac{82861 u_6 u_7 u_2}{45619200 u_1^7}+\frac{373 u_7 u_2}{91238400 u_1}+\frac{650429 u_3 u_8 u_2}{1277337600 u_1^5}+\frac{122413 u_5 u_8 u_2}{91238400 u_1^7}+\frac{41 u_8 u_2}{9580032 u_1^2}\nn\\
 &&+\frac{38377 u_4 u_9 u_2}{53222400 u_1^7}+\frac{3929 u_3 u_{10} u_2}{14515200 u_1^7}+\frac{1091 u_{11} u_2}{174182400 u_1^5}-\frac{19 u_1 u_5 u_2}{13305600}-\frac{23 u_1^2 u_4 u_2}{15966720}-\frac{23 u_3^2 u_2}{21288960} \nn\\
 && -\frac{4801 u_3 u_4 u_2}{766402560 u_1}-\frac{5497 u_4^2 u_2}{63866880 u_1^2}-\frac{26833 u_3 u_5 u_2}{212889600 u_1^2}-\frac{323 u_9 u_2}{68428800 u_1^3}-\frac{10961 u_3 u_6 u_2}{79833600 u_1^3}-\frac{64907 u_4 u_5 u_2}{273715200 u_1^3}\nn\\
 && -\frac{727 u_{10} u_2}{116121600 u_1^4}-\frac{1883363 u_3^2 u_5 u_2}{159667200 u_1^5}-\frac{3705967 u_3 u_4^2 u_2}{239500800 u_1^5}-\frac{7 u_{12} u_2}{4976640 u_1^6}-\frac{14227 u_6^2 u_2}{7096320 u_1^6}-\frac{16217 u_4 u_8 u_2}{7884800 u_1^6}\nn\\
 && -\frac{13001 u_3 u_9 u_2}{15206400 u_1^6}-\frac{864373 u_3^2 u_6 u_2}{53222400 u_1^6}-\frac{1184501 u_4^3 u_2}{106444800 u_1^6}-\frac{542981 u_5 u_7 u_2}{159667200 u_1^6}-\frac{32913731 u_3 u_4 u_5 u_2}{638668800 u_1^6}\nn\\
 && -\frac{1144789 u_4 u_5^2 u_2}{11612160 u_1^8}-\frac{946477 u_3^2 u_8 u_2}{45619200 u_1^8}-\frac{176093453 u_3^3 u_5 u_2}{159667200 u_1^8}-\frac{454075417 u_3^2 u_4^2 u_2}{212889600 u_1^8}-\frac{28069627 u_3 u_4 u_7 u_2}{319334400 u_1^8}\nn\\
 &&-\frac{80480347 u_3 u_5 u_6 u_2}{638668800 u_1^8}-\frac{103432013 u_4^2 u_6 u_2}{1277337600 u_1^8}-\frac{18599541 u_3^4 u_4 u_2}{1576960 u_1^{10}}+\frac{10673 u_3^5}{691200 u_1^7}+\frac{16649 u_3^4}{91238400 u_1^4}\nn\\
 &&+\frac{275599 u_3 u_4^3}{3379200 u_1^7}+\frac{800453 u_4^3}{1916006400 u_1^4}+\frac{571213 u_5^3}{383201280 u_1^7}+\frac{64482661 u_3^3 u_4^2}{60825600 u_1^9}+\frac{1217 u_4^2}{1149603840}+\frac{2264879 u_3 u_5^2}{1277337600 u_1^5}\nn\\
 &&+\frac{7003 u_5^2}{348364800 u_1^2}+\frac{66653 u_3 u_6^2}{28385280 u_1^7}+\frac{u_1^5 u_3}{15966720}+\frac{35113699 u_3^4 u_5}{85155840 u_1^9}+\frac{2932793 u_4^2 u_5}{1277337600 u_1^5}+\frac{4643 u_3 u_5}{1916006400}\nn\\
 &&+\frac{30303149 u_3^2 u_4 u_5}{159667200 u_1^7}+\frac{18211 u_3 u_4 u_5}{9580032 u_1^4}+\frac{1867 u_4 u_5}{79833600 u_1}+\frac{570989 u_3^3 u_6}{14192640 u_1^7}+\frac{13 u_1^2 u_6}{174182400}+\frac{62701 u_3^2 u_6}{106444800 u_1^4}\nn\\
 &&+\frac{8669 u_3 u_6}{638668800 u_1}+\frac{102911 u_3 u_4 u_6}{35481600 u_1^5}+\frac{20521 u_4 u_6}{638668800 u_1^2}+\frac{104281 u_4 u_5 u_6}{14192640 u_1^7}+\frac{4717 u_3^2 u_7}{6082560 u_1^5}\nn\\
 &&+\frac{155959 u_4^2 u_7}{60825600 u_1^7}+\frac{6911 u_1 u_7}{11496038400}+\frac{29611 u_3 u_7}{1916006400 u_1^2}+\frac{318517 u_3 u_5 u_7}{79833600 u_1^7}+\frac{299 u_6 u_7}{1689600 u_1^5}+\frac{1030877 u_3 u_4 u_8}{425779200 u_1^7}\nn\\
 &&+\frac{31 u_5 u_8}{237600 u_1^5}+\frac{7697 u_3^2 u_9}{15206400 u_1^7}+\frac{44647 u_4 u_9}{638668800 u_1^5}+\frac{71 u_3 u_{10}}{2721600 u_1^5}+\frac{13 u_{10}}{74649600 u_1^2}+\frac{113 u_{11}}{348364800 u_1^3}\nn\\
 && +\frac{u_{13}}{29859840 u_1^5}-\frac{23 u_1^2 u_3^2}{21288960}-\frac{u_1^4 u_4}{130636800}-\frac{2011 u_1 u_3 u_4}{638668800}-\frac{79 u_8}{638668800}-\frac{241 u_1^3 u_5}{958003200}-\frac{u_1^8}{1277337600}\nn\\
 && -\frac{361 u_9}{547430400 u_1}-\frac{5707 u_3^3}{3832012800 u_1}-\frac{2099 u_3^2 u_4}{15966720 u_1^2}-\frac{931 u_4 u_7}{22809600 u_1^3}-\frac{13261 u_3 u_4^2}{54743040 u_1^3}-\frac{10277 u_3^2 u_5}{58060800 u_1^3}\nn\\
 &&-\frac{17527 u_3 u_8}{958003200 u_1^3}-\frac{113089 u_5 u_6}{1916006400 u_1^3}-\frac{u_{12}}{4976640 u_1^4}-\frac{2533 u_3 u_9}{106444800 u_1^4}-\frac{8017 u_6^2}{141926400 u_1^4}-\frac{8189 u_4 u_8}{141926400 u_1^4}\nn\\
 &&-\frac{30553 u_5 u_7}{319334400 u_1^4}-\frac{809239 u_3^3 u_4}{106444800 u_1^5}-\frac{11 u_3 u_{11}}{1741824 u_1^6}-\frac{22349 u_4 u_5^2}{3193344 u_1^6}-\frac{701 u_5 u_9}{18247680 u_1^6}-\frac{2143 u_6 u_8}{36495360 u_1^6}\nn\\
 &&-\frac{775 u_4 u_{10}}{41803776 u_1^6}-\frac{520931 u_3^3 u_5}{42577920 u_1^6}-\frac{244457 u_4^2 u_6}{42577920 u_1^6}-\frac{659977 u_3 u_4 u_7}{106444800 u_1^6}-\frac{3697 u_7^2}{109486080 u_1^6}-\frac{1138241 u_3 u_5 u_6}{127733760 u_1^6}\nn\\
 &&-\frac{185851 u_3^2 u_8}{127733760 u_1^6}-\frac{10135361 u_3^2 u_4^2}{425779200 u_1^6}-\frac{336827 u_3^2 u_4 u_6}{2956800 u_1^8}-\frac{3980637 u_3^4 u_4}{7884800 u_1^8}-\frac{653701 u_4^4}{34214400 u_1^8}\nn\\
 &&-\frac{877403 u_3^3 u_7}{42577920 u_1^8}-\frac{8134913 u_3 u_4^2 u_5}{45619200 u_1^8}-\frac{22151509 u_3^2 u_5^2}{319334400 u_1^8}-\frac{4543 u_3^6}{8192 u_1^{10}}. \label{F5}
\eeqa

\end{document}